# Retractions of Types with Many Atoms

Laurent Regnier
Institut de Mathématiques de Luminy
163 Avenue de Luminy, Case 907
13288 Marseille Cedex 9, France
regnier@iml.univ-mrs.fr

Paweł Urzyczyn*
Uniwersytet Warszawski
Instytut Informatyki
Banacha 2, 02-097 Warsaw, Poland
urzy@mimuw.edu.pl

**Corrected version, September 2002**

**Abstract**

We define a sound and complete proof system for affine $\beta\eta$-retractions in simple types built over many atoms, and we state simple necessary conditions for arbitrary $\beta\eta$-retractions in simple and polymorphic types.

## 1 Introduction

Whenever functions between different domains are of interest, it is always the case that the notion of a monomorphism and an isomorphism are fundamental. If we think of typed lambda calculi as of theories of functions between types understood as domains, we may be surprised by how little is understood about mono- and epi-morphisms between types.

In the untyped lambda calculus, there are no domains, and there is a notion of an *invertible* term, corresponding to an abstract notion of an isomorphism, and the notions of right and left invertibility, corresponding to epi- and monomorphisms, respectively. The main results concerning right and/or left invertibility were obtained mostly in the 70's (see e.g. [2, 4, 9]) and an exposition of the theory can be found in Chapter 21 of [1]. But not everything has been understood in full, and there is still a progress in this line of research, see [6].

The first paper, to our best knowledge, that provides a characterization of isomorphic types (simple types with products and an inhabited type constant), is Soloviev's work [12] from 1981, published in English in 1983. Bruce and Longo have obtained a similar characterization in [3], and they also introduced the notion of a *type retraction*. In fact, they only consider $\beta$-retractions. It turns out however, that the more general case of $\beta\eta$-retractions is more difficult.

We say that a type $\rho$ is a *retract* of of a type $\tau$, and we write $\rho \trianglelefteq \tau$, iff there exists a pair of terms $F : \rho \to \tau$ and $G : \tau \to \rho$ such that $G \circ F =_{\beta\eta} \mathbf{I}$, where composition $G \circ F$ is understood as $\lambda x.G(Fx)$. The subtlety in this definition is that types of free variables in $F$ and $G$ may be arbitrary, so a more adequate statement (the lack of which causes confusion among the readers of [3]) is "$E \vdash F : \rho \to \tau$ and $E \vdash G : \tau \to \rho$, for a certain type environment $E$". If the

*Partly supported by KBN Grant 8 T11C 028 20. An essential part of this work was done while the second author was visiting IML in Marseille, thanks to CNRS funding.

environment $E$ is empty (i.e., $F$ and $G$ are combinators) then we say that $\rho$ is *embedded* into $\tau$, and we write $\rho \leq \tau$.

If we replace $\beta\eta$-equality in the above definition by $\beta$-equality (i.e., when $G \circ F =_\beta \mathbf{I}$) we obtain the notion of a $\beta$-retraction (resp. $\beta$-embedding), denoted by $\rho \trianglelefteq_\beta \tau$ (resp. $\rho \leq_\beta \tau$).

We say that types $\rho$ and $\tau$ are *isomorphic* iff there are combinators $F$ and $G$, satisfying both $G \circ F =_{\beta\eta} \mathbf{I}$ and $F \circ G =_{\beta\eta} \mathbf{I}$. The definition of a $\beta$-isomorphism is similar.

All the above definitions are generic in that they apply to every reasonable type system.

The notions of $\beta$-retractability and $\beta$-embeddability in simple types can be easily characterized as follows (see [8, 13]):

$$\rho \trianglelefteq_\beta \tau \quad \text{iff} \quad \tau = \tau_1 \to \cdots \to \tau_k \to \rho, \text{ for some } \tau_1, \dots, \tau_k;$$

$$\rho \leq_\beta \tau \quad \text{iff} \quad \tau = \tau_1 \to \cdots \to \tau_k \to \rho, \text{ for some } \tau_1, \dots, \tau_k, \text{ such that } \tau \vdash \tau_i, \text{ for all } i.$$

In the above, the notation $\tau \vdash \tau_i$ stands for provability in the intuitionistic propositional logic, or equivalently for the existence of a witness term $M$ with $x : \tau \vdash M : \tau_i$. In particular it follows that two-way retractability implies equality of types.

Thus, the case of simply typed $\beta$-retracts and $\beta$-embeddings is well understood. This is however not as easy for the general $\beta\eta$-case. It is not difficult to see that e.g. $a \to b \to c \leq b \to a \to c$, while $a \to b \to c \not\trianglelefteq_\beta b \to a \to c$. The next two examples

$$a \to c \leq a \to a \to c \quad \text{but} \quad a \to c \not\leq_\beta a \to a \to c;$$

$$a \to a \leq ((a \to a) \to a) \to a \quad \text{but} \quad a \to a \not\trianglelefteq_\beta ((a \to a) \to a) \to a,$$

demonstrate, however, that the difference caused by the presence of $\eta$ is not merely the order of arguments.

The paper [8] by de' Liguoro, Piperno, and Statman proves results on retractability in simple types built over a single (inhabited) atom $o$. In particular, they show that retractability is equivalent to the existence of a surjection in every model, and that two-way retractability implies isomorphism. This result is obtained by a semantic approach via Statman's theorem [14]. Another main result of [8] is a complete proof system for "linear", or more adequately, *affine* retractability, that is retractability by means of BCK-terms (at most one occurrence of each variable).

Padovani, in a recent paper [11], proves that the retractability relation is decidable, still under the single atom assumption. For this, he shows that if $\rho \trianglelefteq \tau$ then the coder-decoder pair $F : \rho \to \tau$ and $G : \tau \to \rho$ can be chosen in a certain syntactic form. A very useful generalization used by Padovani, which made it possible to use structural induction, was to replace the binary relation $\rho \trianglelefteq \tau$ between types by a relation $\Sigma \trianglelefteq \tau$ of simultaneous retractability of a type $\tau$ onto a finite product, or equivalently, a multiset of types $\Sigma$.

The following fact is essential for the proof method in [11]:

$$\text{If } \rho_1 \to \cdots \to \rho_n \to o \trianglelefteq \tau_1 \to \cdots \to \tau_m \to o \text{ then } \forall i \leq n \exists j \leq m (\rho_i \trianglelefteq \tau_j). \qquad (*)$$

Unfortunately, this property does not hold with many atoms as can easily be seen from the following example (cf. Lemma 3.3(2)):

$$b \to a \trianglelefteq ((b \to a) \to a) \to a \quad \text{but} \quad b \not\trianglelefteq (b \to a) \to a.$$

Thus, the proof in [11] does *not* carry over to the general case. The same example shows that with many atoms the axiom A4 of [8] is not correct, in consequence the system LPS of de' Liguoro, Piperno, and Statman is unsound. In this paper we show how to modify the axiom A4 to obtain a version of the LPS system, which is sound and complete for affine retractions with many atoms (Section 3). The proof of completeness is via an auxiliary system, which we call CF. At the cost of more complicated rules, the system CF is cut-free, and thus directly translatable into an algorithm to decide affine retractability.

In the present paper we make a further step towards a solution of the still open problem of a sound and complete proof system for arbitrary retractions, namely we state a simple necessary condition for retractability. We show that $\rho_1 \to \cdots \to \rho_n \to a \trianglelefteq \tau$ implies that each $\rho_i$ is a retract of some part of $\tau$ (cf. Padovani's condition (*)). It follows that under an appropriate tree-like representation of finite types, each labeled path in $\rho$ can be expanded to a path in $\tau$.

In particular it follows that:

$$\text{If} \quad \rho \le \tau \quad \text{then} \quad \mathit{rank}(\rho) \le \mathit{rank}(\tau), \qquad (**)$$

where *rank* is the depth of our (non-standard) tree representation of types. More precisely, $\mathit{rank}(a) = 0$ and $\mathit{rank}(\tau_1 \to \cdots \to \tau_n \to a) = 1 + \max\{\mathit{rank}(\tau_i) \mid i = 1, \ldots, k\}$. Condition (**) was first shown by Padovani [11] for types over a single atom.

A comprehensive study of type isomorphisms in typed lambda calculi is the book [5] of Di Cosmo. He gives complete proof rules for isomorphisms of simple as well as polymorphic types. However, there is no discussion of embeddings nor retractions for polymorphic types.

The notion of a polymorphic embedding was first discussed in [13] in connection with the issue of representability of recursive[1] types in system **F**. In order to show that the type $\mu a(b \to a)$ cannot be properly represented by means of beta-reductions, it is shown that there is no type $\tau$ such that $b \to \tau$ can be embedded into $\tau$. This fact follows from the following necessary condition:

$$\text{If} \quad \rho \le_\beta \tau \quad \text{then} \quad d(\rho) \le d(\tau),$$

where $d(\sigma)$ is the depth of the *ordinary* representation of $\sigma$ as a finite tree. That is, $d(a) = 0$, $d(\forall a.\sigma) = d(\sigma)$ and $d(\sigma \to \sigma') = 1 + \max(d(\sigma), d(\sigma'))$.

It is not difficult to observe that the above necessary condition does not hold for $\beta\eta$-embeddability or retractability. However, for some time it was conjectured that condition (**) above was true also for polymorphic types (if we additionally defined $\mathit{rank}(\forall a.\sigma) = \mathit{rank}(\sigma)$). Example 5.1 gives a surprisingly simple counterexample to this conjecture. Instead, we can show a weaker necessary condition, namely

$$\text{If} \quad \rho \trianglelefteq \tau \quad \text{then} \quad FV(\rho) \subseteq FV(\tau),$$

Although very simple and natural, this property requires the same path computation technique as the result in [13]. In fact, we believe that a further refinement of this technique should give a solution of the recursive types problem. In particular, we conjecture the following:

**Conjecture 1.1** *If $\rho \le \tau$ and $\tau \le \rho$ then $\rho$ and $\tau$ are isomorphic.*

If the above conjecture holds, then $\forall b(b \to \tau) \not\le \tau$, for all types $\tau$, and thus the recursive type $\mu a \forall b(b \to a)$ cannot be defined in System **F**.

[1] Note the difference between "recursive" and "inductive". Inductive types are representable, see e.g. [7, 13].

## 2 Preliminaries

Types are denoted by $\tau, \sigma, \ldots$ and type variables by $a, b, c, \ldots$ Multisets of types are written as $\Sigma, \Delta, \ldots$ If $\tau = \tau_1 \to \cdots \to \tau_k \to a$ then we write $a = \mathit{head}(\tau)$. If $\Sigma = \{\tau_1, \ldots, \tau_k\}$ and $\mathit{head}(\tau_1) = \cdots = \mathit{head}(\tau_k) = a$ then we also write $\mathit{head}(\Sigma) = a$.

The relation $\sim$ on simple types is defined as the least equivalence relation satisfying:

1) $\tau \sim \tau'$ and $\sigma \sim \sigma'$ implies $(\tau \to \sigma) \sim (\tau' \to \sigma')$;

2) $(\tau \to \sigma \to \rho) \sim (\sigma \to \tau \to \rho)$.

Types $\rho$ and $\tau$ are *isomorphic* if there are combinators $F : \rho \to \tau$ and $G : \tau \to \rho$ such that both $F \circ G$ and $G \circ F$ are $\beta\eta$-equal to $\mathbf{I}$. For simple types, it follows from [12, 3, 5] that $\rho$ and $\tau$ are isomorphic if and only if $\rho \sim \tau$.

Types may be represented as finite trees in various ways. The most common method is to think of $\sigma \to \tau$ as of a binary tree, the root of which is labeled by the arrow (or just unlabeled) and the two subtrees represent $\sigma$ and $\tau$. Of course, a variable $a$ is seen as a single node labeled $a$. In case of a polymorphic type beginning with a sequence of quantifiers $\forall\vec{a}$, this sequence of quantifiers is an additional label of the root node.

Another approach is to represent a type $\sigma_1 \to \sigma_2 \to \cdots \to \sigma_n \to a$ as a tree with root labeled $a$ and $n$ direct subtrees representing $\sigma_1, \ldots, \sigma_n$. For example, type $(a \to b) \to (a \to c) \to d \to a$ is associated the following tree:

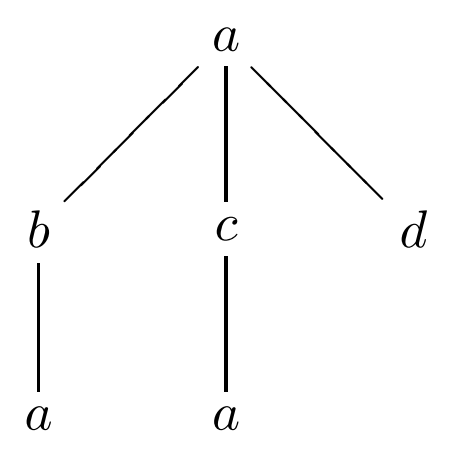

We do not know about the origin of this representation (far less common than the "ordinary" one). We learned it from Hanno Nickau [10].

This idea can be extended to the polymorphic case, provided we agree to identify types $\forall a(\tau \to \sigma)$ and $\tau \to \forall a\sigma$, whenever $a$ is not free in $\tau$. Then a type of the form $\forall\vec{a}_1.\sigma_1 \to \forall\vec{a}_2.\sigma_2 \to \cdots \to \forall\vec{a}_n.\sigma_n \to \forall\vec{a}_{n+1}.a$ may be represented in a similar way, with all labels $\forall\vec{a}_i$ attached at the root. For instance, the type $\forall a.(a \to \forall c(c \to c)) \to \forall b.b \to a$ is represented by

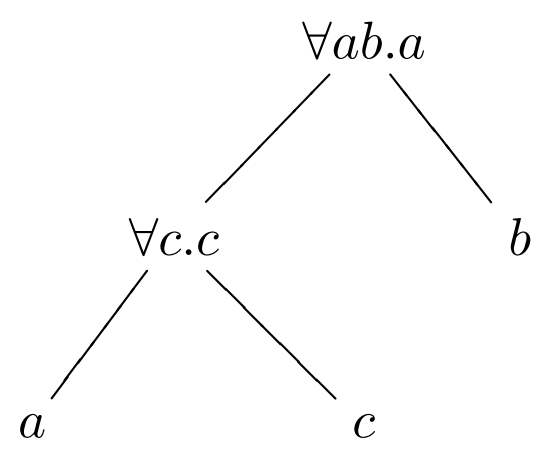

It should be clear that for a simple type $\tau$ the depth of the first representation of $\tau$ is $d(\tau)$, while the depth of the second representation is $\mathit{rank}(\tau)$. We will always refer to the second representation. Note that, in this representation, every node corresponds to a target variable of a different part of $\tau$. Let $\mathit{tree}(\sigma)$ be the tree associated to a type $\sigma$. For every path $\pi$ in $\mathit{tree}(\sigma)$, let $w(\pi)$ be the sequence of type variables occurring on $\pi$, read in the bottom-up direction.

Typed lambda terms are normally assumed to be in Church style, i.e., with full type information available. If a term $M$ is of type $\sigma$, we sometimes write $M^\sigma$ to express this fact, but the superscript is not obligatory.

A type $\rho$ is called a *retract* of a type $\tau$, written $\rho \trianglelefteq \tau$, if there exists a type environment $E$, and terms $C, D$, satisfying the following conditions:

| | |
|---|---|
| 1) $E(x : \rho) \vdash C : \tau$; | 2) $E \vdash D : \tau \to \rho$; |
| 3) $DC =_{\beta\eta} x$; | 4) $x$ is not free in $D$. |

The terms $C$ and $D$ are respectively called the *coder* and the *decoder*. The variable $x$ is called the *main* variable of $C$.

A lambda term $M$ is called *affine* (or a BCK-term) iff each variable occurs in $M$ at most once. Formally, each variable is an affine term, and $\lambda x M$ (respectively $MN$) are affine terms when $M$ is affine (resp. when $M$ and $N$ are affine.)

We write $\rho \trianglelefteq^1 \tau$ iff $\rho \trianglelefteq \tau$ holds with an affine coder and decoder.

The relation $\trianglelefteq$ (and also $\trianglelefteq^1$) is invariant with respect to type isomorphism. Thus, it may be useful to assume in the following that isomorphic types are simply identified. (Note that identifying isomorphic types corresponds to ignoring the order of sons of any given node in a tree representation.) This suggests the following convention: For a multiset of types $\Sigma = \{\tau_1, \ldots, \tau_k\}$, the notation $\Sigma \to \rho$ stands for any representative of the equivalence class $[\tau_1 \to \cdots \to \tau_k \to \rho]_\sim$ (or it is just $\rho$, if $\Sigma$ is empty). Whenever we use this notation, the properties in question are invariant with respect to $\sim$.

It is sometimes convenient to assume that coders and decoders are $\eta$-expanded as much as possible. We define a notion of a *fully $\eta$-expanded* term (simply typed, or polymorphic) as follows:

1. If $N^\sigma$ is fully $\eta$-expanded then so is $(\lambda x{:}\tau.N)^{\tau\to\sigma}$;
2. If $N^\sigma$ is fully $\eta$-expanded then so is $(\Lambda a.N)^{\forall a\sigma}$;
3. If $N$ is fully $\eta$-expanded, $b$ is an atom type, and for some sequence $\vec{\Delta}$ of types and fully $\eta$-expanded terms we have $N\vec{\Delta} : b$, then $N\vec{\Delta}$ is fully $\eta$-expanded.

It should be clear that full $\eta$-expansion is preserved by beta reductions, and that each term has a unique full $\eta$-expansion. A fully $\eta$-expanded normal form is called a *long normal form*.

If $\tau = \{\tau_1, \ldots \tau_n\} \to a$, then we say that each $\tau_i$ is a *part* of $\tau$ *under the path* $a$. Type $\rho$, which is a part of some $\tau_i$ under a path $w$, is said to be a part of $\tau$ under $wa$. A *delayed argument* of $\tau$ is a part of $\tau$ under a path of the form $a^k$ for an odd $k$.

# 3 A system for affine retractions

The following are axioms and rules of the system LPS of de' Liguoro, Piperno and Statman [8]. This system is sound and complete for affine retractions over a single atom $a$.

**Axioms**

(A1) $\sigma \trianglelefteq^1 \sigma$ (A2) $\sigma \trianglelefteq^1 \tau \to \sigma$

(A3) $\sigma \to \rho \to \tau \trianglelefteq^1 \rho \to \sigma \to \tau$ (A4) $\sigma \trianglelefteq^1 (\sigma \to a) \to a$

**Rules**

(R1) $\dfrac{\sigma \trianglelefteq^1 \tau, \quad \tau \trianglelefteq^1 \rho}{\sigma \trianglelefteq^1 \rho}$ (R2) $\dfrac{\sigma \trianglelefteq^1 \sigma', \quad \tau \trianglelefteq^1 \tau'}{\sigma \to \tau \trianglelefteq^1 \sigma' \to \tau'}$

It is not difficult to see that axiom (A4) is not adequate (see Lemma 3.3(2)) unless we add the following restriction:

$$head(\sigma) = a.$$

Of course, now $a$ is an arbitrary atom. Due to our convention of identifying isomorphic types, we may also skip axiom (A3). Thus, from now on, by *system LPS* we mean the system consisting of rules (R1), (R2) and axioms (A1), (A2), (A4), the latter in the modified form:

(A4) $\sigma \trianglelefteq^1 (\sigma \to a) \to a$, provided $head(\sigma) = a$.

We will show the soundness and completeness of system LPS. We find it also convenient to consider a "cut-free" system without the transitivity rule (R1). Our cut-free system is called CF. It has only one axiom and the following rules :

(Axiom) $a \trianglelefteq a$

(H) $\dfrac{\sigma \trianglelefteq^1 \sigma', \quad \tau \trianglelefteq^1 \tau'}{\sigma \to \tau \trianglelefteq^1 \sigma' \to \tau'}$

(N) $\dfrac{\rho \trianglelefteq^1 \tau}{\rho \trianglelefteq^1 \Sigma \to \tau}$

(D) $\dfrac{\Delta_1 \to a \trianglelefteq^1 \sigma_1, \ldots, \Delta_n \to a \trianglelefteq^1 \sigma_n}{\Delta_1 \cup \ldots \cup \Delta_n \to a \trianglelefteq^1 \{\Sigma_1 \to \sigma_1 \to a, \ldots, \Sigma_n \to \sigma_n \to a\} \to a}$

Clearly, rule (H) is he same as (R2). Also, it is not difficult to see that the three rules can be combined into one, namely

(U) $\dfrac{\rho_1 \trianglelefteq^1 \sigma_1, \ldots, \rho_m \trianglelefteq^1 \sigma_m, \ \Delta_1 \to a \trianglelefteq^1 \sigma_1, \ldots, \Delta_n \to a \trianglelefteq^1 \sigma_n}{\{\rho_1, \ldots, \rho_m\} \cup \Delta_1 \cup \ldots \cup \Delta_n \to a \trianglelefteq^1 \Sigma \cup \{\sigma_1, \ldots, \sigma_m\} \cup \{\Sigma_1 \to \sigma_1 \to a, \ldots, \Sigma_n \to \sigma_n \to a\} \to a}$

**Proposition 3.1** *The system LPS is sound for affine retractions: if $\sigma \trianglelefteq^1 \tau$ is derivable then $\sigma \trianglelefteq^1 \tau$ actually holds.*

**Proof:** Similar to that in [8]. ∎

**Proposition 3.2** *If $\sigma \trianglelefteq^1 \tau$ is derivable in system CF then it is also derivable in system LPS. (In particular, CF is sound for affine retractions.)*

**Proof:** Induction with respect to the length of derivations. We consider only the case of rule (D). Let $\Delta_1 \to a \trianglelefteq^1 \sigma_1, \ldots, \Delta_n \to a \trianglelefteq^1 \sigma_n$. First observe that

$$\Delta_n \to a \trianglelefteq^1 ((\Delta_n \to a) \to a) \to a \trianglelefteq^1 (\sigma_n \to a) \to a \trianglelefteq^1 (\Sigma_n \to \sigma_n \to a) \to a$$

The first inequality follows from (A4), the second from (R2) used twice. The last one follows from (A2) and (R2). Since $\Delta_1 \cup \ldots \cup \Delta_n \to a$ is the same as $(\Delta_1 \cup \ldots \cup \Delta_{n-1}) \to \Delta_n \to a$, it follows that

$$\Delta_1 \cup \ldots \cup \Delta_n \to a \trianglelefteq^1 (\Delta_1 \cup \ldots \cup \Delta_{n-1}) \to (\Sigma_n \to \sigma_n \to a) \to a.$$

But the latter type is isomorphic to $\{\Sigma_n \to \sigma_n \to a\} \cup \Delta_1 \cup \ldots \cup \Delta_{n-1} \to a$, and we can use a similar reasoning for $\Delta_{n-1}$, obtaining

$$\{\Sigma_n \to \sigma_n \to a\} \cup \Delta_1 \cup \ldots \cup \Delta_{n-1} \to a \trianglelefteq_1 (\{\Sigma_n \to \sigma_n \to a\} \cup \Delta_1 \cup \ldots \cup \Delta_{n-2}) \to (\Sigma_{n-1} \to \sigma_{n-1} \to a) \to a.$$

After repeating this argument $n$ times one ends up with the conclusion of the rule. ∎

**Lemma 3.3**

1) *If $D^{\tau\to\rho} C^\tau =_{\beta\eta} x^\rho$, and $x \notin FV(D)$ then the long normal form of $D$ is of shape $\lambda f^\tau.\lambda \vec{z}.f\vec{M}$.*
2) *If $\rho \trianglelefteq \tau$ then $head(\rho) = head(\tau)$.*

**Proof:** Part (1) is easy. Use (1) to prove (2). ∎

**Lemma 3.4** *Let $M = M^0[X := xV_1 \ldots V_k] \twoheadrightarrow xA_1 \ldots A_k B_1 \ldots B_r$, where $M$ is an affine term. Then $M^0 \twoheadrightarrow XB_1 \ldots B_r$.*

**Proof:** The proof is by induction with respect to the standard (leftmost) reduction. First suppose that $M^0$ has a head redex: $M^0 = (\lambda w.N)P\vec{Q} \longrightarrow N[w := P]\vec{Q}$. Then $M = ((\lambda w.N)P\vec{Q})[X := xV_1 \ldots V_k] \longrightarrow (N[w := P]\vec{Q})[X := xV_1 \ldots V_k]$, and we can apply the induction hypothesis for $(N[w := P]\vec{Q})$.

If $M^0$ does not have a head redex, then $M^0$ must already in the form $XB'_1 \ldots B'_r$, where $B'_i \twoheadrightarrow B_i$, for all $i = 1, \ldots, r$. ∎

**Theorem 3.5** *The system CF is complete for affine retractions: every true statement $\rho \trianglelefteq^1 \tau$ is derivable.*

**Proof:** The proof is by induction with respect to the size of $\rho \trianglelefteq^1 \tau$. Let $\rho = \rho_1 \to \cdots \to \rho_\ell \to a$ and let $\tau = \tau_1 \to \cdots \to \tau_n \to a$. If $\ell = 0$, i.e., $\rho = a$ then $\rho \trianglelefteq^1 \tau$ is easily derived from the axiom by applying rule (N) as many times as needed. Let $\ell > 0$ and assume that the coder and the decoder are given in long normal forms. We have $C \equiv \lambda y_1^{\tau_1} \ldots y_n^{\tau_n}.C'$, and the decoder $D$ must be of the form $\lambda f^\tau \lambda z_1^{\rho_1} \ldots z_\ell^{\rho_\ell}.fS_1^{\tau_1} \ldots S_n^{\tau_n}$ (cf. Lemma 3.3(1)). There is exactly one occurrence of each $z_i$ in $D$, within some $S_j$. We use the notation $j = t(i)$. Note that, due to affinity, the variable $f$ does not occur free in the $S_j$'s.

Each $j = 1, \dots, n$ is classified as either *passive* or *active*, as follows. Take a fresh variable $*$ of the same type as $y_j$ and consider the term $C_j^* = \lambda y_1^{\tau_1} \dots y_n^{\tau_n}.C'[y_j := *]$. Reduce the application $DC_j^*$ to a head normal form, and inspect the head variable $u$. It should be clear that it must be either $x$ or $*$, as otherwise $u$ would be equal to the head variable of $DC$, that is to $x$. If $u = x$ then we classify $j$ as *passive*, otherwise $j$ is *active*.

Consider any $i = 1, \dots, \ell$. We now proceed by cases depending on $j = t(i)$.

**Case 1:** Assume that $j$ is passive. Consider the standard reduction of $DC_j^*$ to its head normal form:

$$DC_j^* \twoheadrightarrow xN_1^{\rho_1} \dots N_\ell^{\rho_\ell},$$

and reduce the term $DC$ according to the same pattern. We have a reduction of the following form:

$$\begin{aligned} DC &\twoheadrightarrow_\beta \lambda z_1 \dots z_\ell.(\lambda y_1^{\tau_1} \dots y_n^{\tau_n}.C')S_1^{\tau_1} \dots S_n^{\tau_n} \twoheadrightarrow_\beta \lambda z_1 \dots z_\ell.C'[y_r := S_r]_{r=1}^n \twoheadrightarrow_\beta \\ &\twoheadrightarrow_\beta \lambda z_1 \dots z_\ell.xM_1^{\rho_1} \dots M_\ell^{\rho_\ell} \end{aligned}$$

This differs from the reduction of $DC_j^*$ in that we have $S_j$ in place of $*$. The term $S_j$ does not get substituted on the way, because there is no variable free in $S_j$ that would be bound outside of it. Since $z_i$ must occur in $M_i$, we have that $M_i = N_i[* := S_j]$. (Note that this means in particular that $t(j) \neq t(j')$ whenever $j \neq j'$.) In addition, $M_i$ must reduce to $z_i$. Now we can conclude that

$$\rho_i \trianglelefteq^1 \tau_j.$$

Indeed, we can take $S_j$ as the coder (with main variable $z_i$), and $\lambda *.N_i$ as the decoder.

**Case 2:** Now let $j$ be active. With no loss of generality, we may assume that $t^{-1}(j) = \{k, \dots, \ell\}$ (a final segment of arguments). Assume that $\tau_j = \sigma_1 \to \dots \to \sigma_p \to a$. Again we "simulate" the head reduction of $DC_j^*$ and we obtain

$$\begin{aligned} DC &\twoheadrightarrow_\beta \lambda z_1 \dots z_\ell.(\lambda y_1^{\tau_1} \dots y_n^{\tau_n}.C')S_1^{\tau_1} \dots S_n^{\tau_n} \twoheadrightarrow_\beta \lambda z_1 \dots z_\ell.C'[y_r := S_r]_{r=1}^n \twoheadrightarrow_\beta \\ &\twoheadrightarrow_\beta \lambda z_1 \dots z_\ell.S_j M_1^{\sigma_1} \dots M_p^{\sigma_p} \twoheadrightarrow_\beta \lambda z_1 \dots z_\ell.xz_1 \dots z_\ell. \end{aligned}$$

There is one occurrence of $x$ in the above term. Assume for simplicity that this occurrence is within $M_1$, in a context of the form $xW_1^{\rho_1} \dots W_\ell^{\rho_\ell}$. We claim that

$$\{\rho_k, \dots, \rho_\ell\} \to a \trianglelefteq^1 \sigma_1.$$

As the coder we take the term $M_1'$ obtained from $M_1$ by replacing $xW_1^{\rho_1} \dots W_\ell^{\rho_\ell}$ by $XW_1^{\rho_k} \dots W_\ell^{\rho_\ell}$, where $X$, the main variable, is fresh. The decoder is $E \equiv \lambda g \lambda z_k \dots z_\ell.S_j g M_2^{\sigma_2} \dots M_p^{\sigma_p}$.

We have $EM_1' \longrightarrow_\beta \lambda z_k \dots z_\ell.S_j M_1' M_1^{\sigma_1} \dots M_p^{\sigma_p}$. By Lemma 3.4, the latter term reduces to $\lambda z_k \dots z_\ell.Xz_k \dots z_\ell$.

The conclusion from the two cases discussed above is the following.

a) If $j = t(i)$ is passive then $\rho_i \trianglelefteq^1 \tau_j$, and the "passive part" of the function $t$ is one to one.

b) If $j$ is active then $\{\rho_i \mid t(i) = j\} \to a \trianglelefteq^1 \sigma$, where $\sigma$ is a certain argument of $\tau_j$.

We take types up to isomorphisms, so we can freely assume that $t(i)$ are passive for an initial segment of $\{1, \ldots, \ell\}$ and active otherwise. We can also assume that $j = t(j)$ for the active $j$'s. The other arguments of $\rho$ can be split into the sets $t^{-1}(j)$ for the active $j$'s. From the discussion above it follows that $\rho \trianglelefteq^1 \tau$ is derived from inequalities of the form (a) and (b) with help of rule (U). Indeed, each type of the form $\Sigma_r \to \sigma_r \to a$ is a type of some $y_j$, where $j$ is active and $\sigma$ is the distinguished argument of $\tau_j$. The set $\Sigma$ corresponds to the $j$'s outside the range of $t$ ∎

Recall that the *rank* of a type is defined by $rank(a) = 0$ and $rank(\tau_1 \to \cdots \to \tau_n \to a) = 1 + \max\{rank(\tau_i) : i = 1, \ldots, k\}$. Using Theorem 3.5, one can show by an easy induction that $\rho \trianglelefteq^1 \tau$ implies $rank(\rho) \leq rank(\tau)$. This also holds for non-affine retractions (Proposition 4.5).

Other consequences of Theorem 3.5 are as follows:

**Corollary 3.6**

1. *The relation $\trianglelefteq^1$ is decidable (in nondeterministic polynomial time).*
2. *If $\rho \trianglelefteq^1 \tau$ and $\tau \trianglelefteq^1 \rho$ then $\rho$ and $\tau$ are isomorphic.*

**Proof:** (1) System CF gives a nondeterministic recursive algorithm to verify whether $\rho \trianglelefteq^1 \tau$ holds. The depth of recursive calls is linear, because each time disjoint parts of $\rho$ and $\tau$ are processed.[2]

(2) First we prove that there is no nontrivial retraction of a type onto itself, i.e., that each such retraction is actually an isomorphism. The proof is by induction with respect to the size of types. Clearly, there is no other coder from $a$ to $a$ but identity. For complex types, either rule (H) or (D) must be applied. In case of (H), we can immediately apply the induction hypothesis. Rule (D) is not applicable. Indeed, it follows from the induction hypothesis that the size of the right hand side must be larger than that of the left-hand side.

Now suppose $\rho \trianglelefteq^1 \tau$ with coder $C$ and decoder $D$. Also suppose $\tau \trianglelefteq^1 \rho$ with coder $C'$ and decoder $D'$. Let $F = \lambda x.C$ and $F' = \lambda x'.C'$, where $x$ and $x'$ are the respective main variables. Then $(F \circ F')x$ and $D' \circ D$ make a coder/decoder pair for $\tau \trianglelefteq^1 \tau$. Thus, $F \circ F'$ must be an isomorphism, and we have $F \circ F' \circ D' \circ D = \mathbf{I}$. It follows that $D = D \circ \mathbf{I} = D \circ F \circ F' \circ D' \circ D = \mathbf{I} \circ F' \circ D' \circ D = F' \circ D' \circ D$. But the equality $D = F' \circ D' \circ D$ implies that $F \circ D = F \circ F' \circ D' \circ D = \mathbf{I}$, i.e., that $F$ and $D$ define an isomorphism between $\rho$ and $\tau$. ∎

# 4 Arbitrary retractions in finite types

The proof rules of the previous section are no longer complete if non-affine coders and decoders are permitted. An example was given in [8]. Another one is as follows.

**Example 4.1** Let $\rho = (e \to a) \to c \to a$ and $\tau = (e \to (a \to c \to a) \to a) \to a$.

Then $\rho \trianglelefteq \tau$ with the following coder and decoder:

$$C = \lambda y.yE^e(\lambda w_1^a w_2^c.x(\lambda v^e.yv(\lambda w_1^a w_2^c.w_1))w_2);$$

$$D = \lambda f \lambda z_1^{e\to a} z_2^c.f(\lambda u_1^e u_2^{a\to c\to a}.u_2(z_1 u_1)z_2),$$

[2] It is unclear if the nondeterminism can be eliminated. An exponential choice may be caused by rule (D).

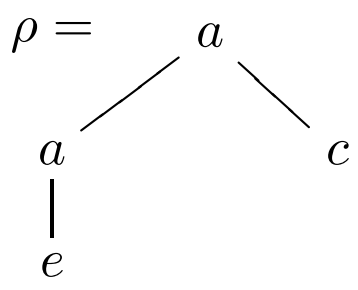

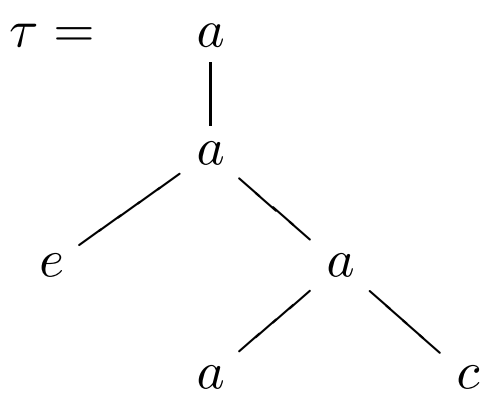

where the variable $y$ (used twice) has type $e \to (a \to c \to a) \to a$, and $E$ is anything of type $e$. But $\rho \not\trianglelefteq^1 \tau$. Indeed, otherwise it is derivable in our system. The last rule in the derivation can't be (N), because $\tau$ has only one argument and $\rho \not\trianglelefteq^1 a$. Similarly, if (H) was the last rule then either $c \to a \trianglelefteq^1 a$ or $(e \to a) \to a \trianglelefteq^1 a$. If the last rule was (D) then either $\rho \trianglelefteq^1 a$ or $\rho \trianglelefteq^1 c \to a \to a$.

**Example 4.2** In the above example (as the one in [8]), $\rho$ is a retract of $\tau$, because $\rho \sim \Sigma \to \rho'$, where all types in $\Sigma$ are retracts of arguments of $\tau$, and the final part $\rho'$ is a retract of an argument of an argument of $\tau$ (corresponding to the head variable of the coder). That this pattern does not make a sufficient condition can be seen when we replace $\tau$ in the above example by $\tau' = (e \to (c \to a) \to a) \to a$.

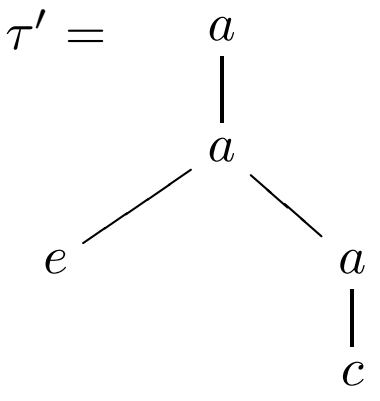

Suppose that $\rho \trianglelefteq \tau'$. Then the long normal form of the decoder must be:

$$D = \lambda f \lambda z_1^{e\to a} z_2^c . f(\lambda u_1^e v_1^{c\to a}. f(\lambda u_2^e v_2^{c\to a} \dots f(\lambda u_k^e v_k^{c\to a}. A^a)\cdots)),$$

for some $k \geq 1$ and for some term $A$ of type $a$, such that the head variable of $A$ is no longer $f$. It should be clear that $A$ must contain both $z_1$ and $z_2$ free, and that the head variable of $A$ must not be free in $D$. But this is impossible. Indeed, in order to be of type $a$, the term $A$ must either be of the form $z_1 u_i$ or of the form $v_i z_2$.

**Example 4.3** This (affine) example shows that the pattern mentioned in Example 4.2 is not obligatory. We have $\rho \trianglelefteq \tau$, where $\rho = b \to c \to a$ and $\tau = ((b \to a) \to a) \to ((c \to a) \to a) \to a$.

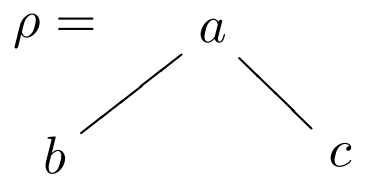

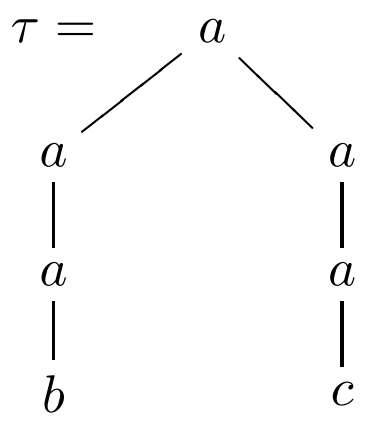

The coder and decoder are as follows:

$$C = \lambda y_1 y_2 . y_1(\lambda w^b . y_2(\lambda u^c . xwu));$$

$$D = \lambda f \lambda z_1^b z_2^c . f(\lambda s^{b \to a} . sz_1)(\lambda r^{c \to a} . rz_2),$$

where $y_1 : (b \to a) \to a$ and $y_2 : (c \to a) \to a$.

Example 4.1 suggests that it may indeed be difficult to design a proof system for the relation $\trianglelefteq$, as it is not clear if $\rho \trianglelefteq \tau$ can be at all expressed in terms of statements of the form $\rho' \trianglelefteq \tau'$, where $\rho'$ and $\tau'$ are, respectively, parts of $\rho$ and $\tau$. On the positive side, we can show the following necessary condition.

**Theorem 4.4** *Let $\rho \trianglelefteq \tau$. Then $head(\rho) = head(\tau)$ and each argument of $\rho$ is a retract of a delayed argument of $\tau$.*

**Proof:** That $head(\rho) = head(\tau) = a$, we already know from Lemma 3.3(2). Assume that $\rho = \rho_1 \to \cdots \to \rho_\ell \to a$ and $\tau = \tau_1 \to \cdots \to \tau_n \to a$. Also assume that the coder $C = \lambda y_1 \ldots y_n . body(C)$ and the decoder $D = \lambda f \lambda z_1 \ldots z_\ell . body(D)$ are given in long normal forms.

By a *segment* we mean a maximal fully $\eta$-expanded subterm of $D$ or of $C$. Each segment has the form $S = \lambda \vec{u} . v \vec{R}$, where the terms $\vec{R}$ are segments. We write $body(S)$ for $v\vec{R}$. The variable $v$ is the *head variable* and the $\vec{u}$ are *leading variables* of the segment.

We will use a sequence of segments to represent an initial fragment of the reduction of the application $DC$. More precisely, we define a sequence $S_1, S_2, \ldots$ of segments, called the *primary spine*, together with an assignment of *meanings* to the leading variables. Variables $z_1, \ldots, z_\ell$ make an exception and have no meanings. Variables occurring free in a segment, but bound in $C$ or $D$, are assigned meanings, because they occur as leading variables earlier on the spine.

First, take $S_1$ to be $D$. Then suppose $S_1, S_2, \ldots, S_k$ have been defined. Take $S_{k+1}$ to be the meaning of the head variable of $S_k$ (which must be part of some $S_q$, with $q < k$). The meaning of the $i$-th leading variable of $S_{k+1}$ is defined as the $i$-th argument in $S_k$. Note that since segments are fully expanded, the number of lambdas and arguments must match. Free variables of $S_{k+1}$ are associated to the same lambda bindings as in $S_q$.

The *final meaning* of a variable $x$ is obtained from its meaning $S$ by replacing free variables of $S$ by their final meanings. Note that a final meaning of a variable does not contain free variables possessing meanings. The *final meaning* of a subterm $M$ of a segment, denoted $[\![M]\!]$, is obtained by replacing free variables of $M$ (whenever possible) by their final meanings.

Now observe that the segments on a spine must alternate: odd numbered segments are parts of $D$, while the even-numbered segments are parts of $C$. This is equivalent to saying that a meaning of a variable from $C$ is always a part of $D$ and conversely. Indeed, the meaning of a leading variable is always taken from the previous segment, and we can prove this property by induction.

Also by easy induction we can show that a free variable of any final meaning must be a free variable of $C$ or of $D$.

The next observation is that

$$[\![body(S_k)]\!] \twoheadrightarrow_\beta [\![body(S_{k+1})]\!].$$

This is because $[\![body(S_k)]\!] = [\![S_{k+1}]\!]\vec{A}$, for some final meanings $\vec{A}$. By performing reduction these final meanings are replacing the leading variables in $body(S_{k+1})$ resulting in $[\![body(S_{k+1})]\!]$.

Since $[\![body(S_1)]\!] = [\![body(D)]\!] = body(D)[f := C]$, we have:

$$DC \twoheadrightarrow_\beta \lambda z_1 \dots z_\ell [\![ \mathit{body}(S_k) ]\!], \text{ for all } k.$$

It follows from strong normalization that

$$[\![ \mathit{body}(S_k) ]\!] \twoheadrightarrow xz_1 \dots z_\ell.$$

In particular, the sequence $S_1, S_2, \dots$ must end exactly when the head variable of some $S_k$ is $x$. Then $[\![ \mathit{body}(S_m) ]\!] = xZ_1 \dots Z_\ell$, for some $Z_1, \dots, Z_\ell$.

Let us now fix a $j \leq \ell$. We define the *secondary spine* $T_1, T_2, \dots$, beginning with $T_1 = Z_j = \lambda v_1 \dots v_r.\mathit{body}(Z_j)$, in the same way as the primary spine, but taking into account meanings of variables defined before. An argument similar to the above shows that

$$[\![ Z_j ]\!] \twoheadrightarrow_\beta \lambda v_1 \dots v_r.[\![ \mathit{body}(T_k) ]\!] \twoheadrightarrow \lambda v_1 \dots v_r.z_j v_1 \dots v_r, \text{ for all } k,$$

and that the secondary spine ends at some $T_q$ which has $z_j$ as the head variable.

The occurrence of $z_j$ as the head variable of $T_q$ is considered *essential*. Further, suppose that an occurrence of $z_j$ in some $T_k$ for $k \leq q$ is essential. If $T_k$ was obtained as a meaning of a variable bound on the secondary spine, and this meaning was taken as a subterm of some $T_s$, for $s < k$, then the appropriate occurrence of $z_j$ in $T_s$ is also essential. Note that occurrences of $z_j$, which are not essential, might be replaced by fresh variables without affecting the reduction.

Let $T_p$ be the first segment on the secondary spine, which contains an essential occurrence of $z_j$. Of course, $T_p$ is part of $D$, so $p$ must be even, in particular $p > 1$. Thus $T_p$ is a meaning of a variable $y$ which is bound in $C$.

We claim the following: if $y$ is a bound variable in $C$ which has a meaning, then the type of $y$ is a part of $\tau$. In addition, this part must be at a distance of an odd length.

This is clear if $y$ is a leading variable of $C$. Otherwise, $y$ occurs as a leading variable of a term $\lambda\vec{y}.A$, which is an argument of another variable $y'$. In order for $y$ to have a meaning, also $y'$ must have a meaning, and thus we can proceed by induction. (Note that the difference in the length of the distances is 2.)

We conclude that the type $\tau'$ of of $T_p$ is a part of $\tau$ at a distance of an odd length. (We still need to know that this distance consists exclusively of $a$'s.) We will show that $\rho_j \trianglelefteq \tau'$ with the coder $[\![T_p]\!]$.

First of all, note that the variable $y$, of which $T_p$ is the meaning must be bound on the primary spine and not on the secondary one. Otherwise, $T_p$ would not be the first segment with an essential occurrence of $z_j$. This also means that variables $v_1 \dots v_r$ do not occur in $[\![T_p]\!]$. Indeed, by an easy induction one can show that a free variable of any final meaning on the primary spine must be a free variable of $C$ or of $D$.

Let the term $T'$ be obtained from $\mathit{body}(T_{p-1})$ by replacing the head variable $y$ by a fresh variable $Y$. Define $T'' = \lambda Y \lambda v_1 \dots v_r.[\![T']\!]$, and let $T$ be $T'$ with all occurrences of $z_j$ replaced by fresh $z'_j$. These occurrences of $z_j$ are not essential, and as we have observed, they have no influence on the reduction. Thus, $T[\![T_p]\!]$ reduces to $\lambda v_1 \dots v_r.z_j v_1 \dots v_r$, i.e., the term $T$ is a correct decoder.

It remains to prove that $\tau'$ is a delayed argument of $\tau$. First observe that the target variable of types of all segments at the primary spine must be $a$, because this must be the same variable as the target of $\rho$, the type of $x$. Let $y$ be the variable of which $T_p$ is the meaning. We have already observed that it is a leading variable of some $S_i$ on the primary spine. If this $S_i$ is the meaning of some $y'$, then the target of the type of $y'$ is $a$, and $y'$ must be bound on the primary spine too. Thus, $\tau'$ is a part of the type of $y'$ at the distance $a$. Note that $y'$ is a variable bound in $D$. Moving this way upward the primary spine, we see that $\tau'$ must be a part of the type $\tau$ of $f$ at a distance built only from $a$'s. ∎

For two words, $w$ and $v$, we write $w \sqsubseteq^0 v$ iff $w = w'aw''$ and $v = w'aaaw''$ for some letter $a$ and some $w', w''$. The symbol $\sqsubseteq$ stands for the transitive and reflexive closure of $\sqsubseteq^0$. The following is an easy consequence of Theorem 4.4.

**Proposition 4.5**

1. *If* $\rho \trianglelefteq \tau$ *then* $rank(\rho) \leq rank(\tau)$.
2. *If* $\rho \trianglelefteq \tau$ *and* $\pi$ *is a path in* $\rho$ *then there is a path* $\pi'$ *in* $\tau$ *with* $w(\pi) \sqsubseteq w(\pi')$.

Unfortunately, Theorem 4.4 does not suffice as a basis for a decision procedure for the relation $\trianglelefteq$, and the decidability of $\trianglelefteq$ remains an open problem. However, the other part of Corollary 3.6 is true, even if we can not give a direct syntactic proof. Indeed, we can modify the semantic proof given in [8] for the single atom case.

**Proposition 4.6** *If* $\rho \trianglelefteq \tau$ *and* $\tau \trianglelefteq \rho$ *then* $\tau$ *and* $\rho$ *are isomorphic.*

**Proof:** It is shown in [8] (Proposition 2.4) that this property holds for types over a single atom. Let $\rho^o$ (resp. $\tau^o$) be obtained from $\rho$ (resp. $\tau$) by replacing all type variables by a single type constant $o$. Of course $\rho \trianglelefteq \tau$ and $\tau \trianglelefteq \rho$ implies that $\rho^o \trianglelefteq \tau^o$ and $\tau^o \trianglelefteq \rho^o$. Proposition 2.4 in [8] states that $\rho^o$ and $\tau^o$ must be isomorphic. It is of course not enough, but what is actually shown is that in this case an arbitrary retraction is an isomorphism. Let $C$ and $D$ be respectively the coder and the decoder for $\rho \trianglelefteq \tau$. Then, of course, $C$ and $D$ work also for $\rho^o \trianglelefteq \tau^o$, and thus $(\lambda x.C) \circ D =_{\beta\eta} \mathbf{I}$. This property does not depend on types assigned to $C$ and $D$, and thus $\tau$ and $\rho$ are isomorphic too. ∎

# 5 The polymorphic case

**Example 5.1** Let $\xi = ((g \to f) \to e) \to d$. Take $\rho = ((\xi \to \xi \to c) \to b) \to a$, and $\tau = \forall\alpha(((\alpha \to \alpha \to c) \to b) \to (\alpha \to \xi) \to a)$. (See the picture.)

Then both the rank (depth) of $\rho$ and its number of nodes are greater than of $\tau$. But we have $\rho \leq \tau$ with:

$$C = \Lambda\alpha\lambda yz.x(\lambda x_2.y(\lambda vu.x_2(zv)(zu))),$$

$$D = \lambda f\lambda x_1.f\xi x_1\mathbf{I}.$$

The above example shows that the properties of $\trianglelefteq$ do not carry over from the simply typed to the polymorphic case. We can still however give the following necessary condition for for polymorphic retractability: $\sigma \trianglelefteq \tau$ implies $FV(\sigma) \subseteq FV(\tau)$. The remainder of this section is devoted to a sketch of proof of this property.

If we assume that coders and decoders are always fully $\eta$-expanded, then a reduction $DC \twoheadrightarrow_{\beta\eta} x$ can be split as $DC \twoheadrightarrow_{\beta} X \twoheadrightarrow_{\eta} x$, where $X$ is the full $\eta$-expansion of $x$. (This is because no $\eta$ redex can block a $\beta$ reduction step.) Thus, from now on, we think of $X$ rather than of $x$, as of the final stage of the reduction, and we only need to talk about $\beta$-reductions.

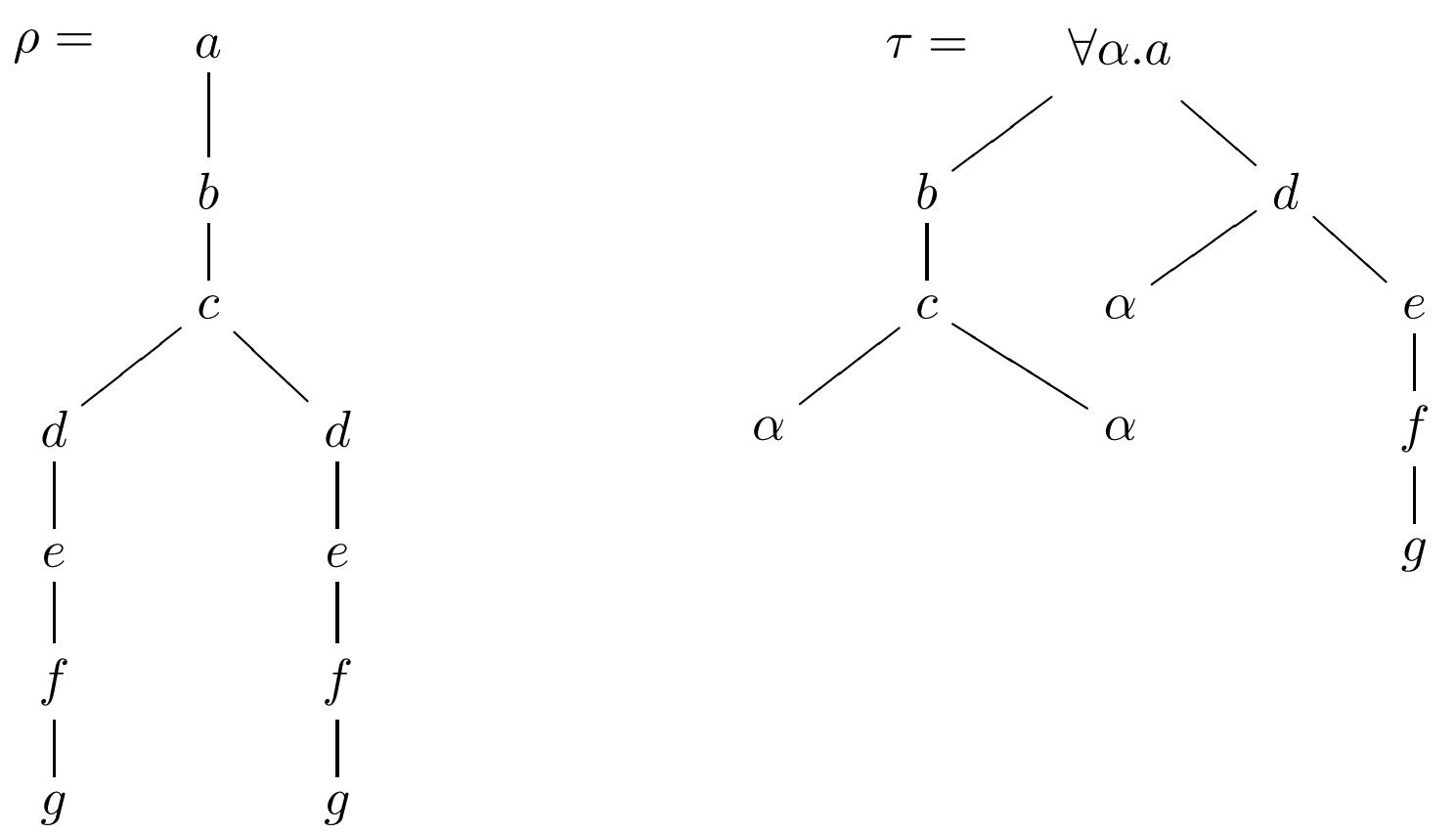

Let us have a closer look at a full $\eta$-expansion $X$ of a variable $x : \rho$. The only free variable of $X$ is $x$ and we may assume that all bound variables are chosen different. It is not difficult to see that there is a bijective correspondence between variables in $X$ and nodes in the tree representation of $\rho$. The variable $x$ corresponds to the root, and then we have: $X = \Lambda\vec{a}\lambda\vec{x}_1.x\vec{a}\vec{X}_1$, where $\vec{X}_1$ are full $\eta$-expansions of $\vec{x}_1$. Each of variables in $\vec{x}_1$ corresponds to a son of the root, and then we apply induction. Write *node*$(y)$ for the node corresponding to $y$. Note that if $y : \rho'$ occurs in $X$ then the target variable of $\rho'$ is the label of *node*$(y)$. The *rank* of $y$ is defined as the depth of *node*$(y)$, and if *node*$(y)$ is a son (father) of *node*$(z)$, then we also refer to $y$ as a son (father) of $z$.

From now on assume that $x : \rho \vdash C : \tau$, $\vdash D : \tau \to \rho$ and $DC \twoheadrightarrow_\beta X$, where $X$ is the full $\eta$-expansion of $x$. Of course $C$ and $D$ are fully $\eta$-expanded. Identify the variables of $X$ with their ancestors in $DC$.

**Lemma 5.2** *(The ancestors of) all variables of $X$ of even depth occur in $C$, while (the ancestors of) all variables of odd depth occur in $D$.*

**Proof:** Of course the ancestor of $x$ occurs in $C$, and it is not difficult to see that $D = \Lambda\vec{a}\lambda f\lambda\vec{x}_1\ldots$. where $\vec{x}_1$ are all the (ancestors of) variables of depth 1. Then we can proceed by induction. A variable $y$ of depth $i$ must occur in $X$ in a context of the form: $\ldots\Lambda\vec{\beta}\lambda\vec{y}_1.y\vec{\beta}\vec{Y}_1$, where $\vec{Y}_1$ are full $\eta$-expansions of $\vec{y}_1$. If some $y_1 \in \vec{y}_1$ is of type of depth greater than zero, then the corrsponding member of $\vec{Y}_1$ is $Y_1 = \Lambda\vec{\gamma}\lambda\vec{y}_2.y_1\vec{Y}_2$. Suppose the ancestor of $y$ occurs in $C$. It must occur in a subterm of the form $y\vec{\delta}\vec{Y}_1'$ which is eventually turned into $y\vec{\beta}\vec{Y}_1$ by a series of substitution. Since $\vec{Y}_1'$ are fully $\eta$-expanded, they must begin with $\Lambda\vec{\gamma}\lambda\vec{y}_2.y_1\ldots$, and thus all variables of level $i+2$ must also occur in $C$. ∎

We represent lambda terms as graphs, and we consider computations along these paths, as in [13]. We refer the reader to the latter paper for detailed definitions.

**Proposition 5.3** *If $\rho \trianglelefteq \tau$ then $FV(\rho) \subseteq FV(\tau)$.*

**Proof:** Let $a \in FV(\rho)$. If $a$ is the target of $\rho$, then it must also be the target of $\tau$, because the decoder must have the form $\Lambda\vec{a}\lambda f.f\ldots$

Otherwise, let $y$ be a variable in $X$, of level $i+1>0$, which corresponds to an occurrence of $a$, and let $z$ be the father of $y$. There is a path $\Pi$ in $X$ from $y$ to $z$ and a computation along $\Pi$, beginning at the configuration $(y,(q^k))$ and ending at $(z,(q^\ell pq^k))$. This path $\Pi$ is obtained as a result of a sequence of deformations from a path $\Pi_0$ in $DC$, also from $y$ to $z$. Using a reasoning almost identical to that of [13], we can show that there is a correct computation along $\Pi_0$, and that this computation is "good", i.e., every compression begins a $\forall$-cycle ending at the same node with a matching decompression. Indeed, the proof of the crucial Lemma 11, remains essentially the same. The only difference is as follows: the reason we must leave a term of the form $(\Lambda a.A)^{\forall a\,\tau}\sigma$ entered via its top node with a compression is not because we must eventually visit the top node of the whole term. Instead, we must reach $z$, with a push-down store pointing to a free occurrence of $a$. However, within the term $\Lambda a.A$ our wires are "below" occurrences of $a$ in all the types they cross.

By Lemma 5.2, the path $\Pi_0$ either begins within $C$ and ends within $D$ or conversely. In each case, we must cross the type $\tau$ of $C$. But if we do it in a compressed configuration, we must necessarily return back. Thus, at least once, we do it without any compression, and this means that $a$ must be free in $\tau$. ∎

Proposition 5.3 is of course useless when e.g. $\rho \trianglelefteq \tau$ and $\rho$ is a closed type. We conjecture that there is always an injection from bound variables of $\rho$ to bound variables of $\tau$. However, a generalization of Proposition 5.3 to handle bound variables will require a finer analysis.

Indeed, in the $\beta\eta$ case, it is in general not true that every compression must have a matching decompression. Consider, for instance a variable $x:\forall a(a\to a)$, and its full $\eta$-expansion $X = \Lambda a\lambda y^a.xay$. Then the path from $y$ to $x$ in $X$ gets "compressed" at the type application $xa$ and is never "decompressed". This happens because $a$ is a bound variable.

The analogue of Proposition 4.6 for polymorphic types is only conjectured.

**Conjecture 5.4** *If $\rho \trianglelefteq \tau$ and $\tau \trianglelefteq \rho$ then $\rho$ and $\tau$ are isomorphic.*

A possible way to prove this conjecture would be to show first that every self-retraction is an isomorphism:

**Conjecture 5.5** *If $\tau \trianglelefteq \tau$ with coder $C$ and decoder $D$ then $(\lambda x.C)\circ D =_{\beta\eta} \mathbf{I}$.*

**Proposition 5.6** *Conjecture 5.5 implies Conjecture 5.4.*

**Proof:** Same as the proof of Corollary 3.6(2) ∎

# Acknowledgement

Thanks are due to Aleksy Schubert for spotting a bug in an earlier version of this paper.